# Absence of the Ordinary and Extraordinary Hall effects scaling in granular ferromagnets at metal-insulator transition.

D. Bartov, A. Segal, M. Karpovski, and A. Gerber

Raymond and Beverly Sackler Faculty of Exact Sciences

School of Physics and Astronomy

Tel Aviv University,

Ramat Aviv 69978 Tel Aviv, Israel

Universality of the extraordinary Hall effect scaling was tested in granular three-dimensional Ni-SiO$_2$ films across the metal-insulator transition. Three types of magnetotransport behavior have been identified: metallic, weakly insulating and strongly insulating. Scaling between both the ordinary and extraordinary Hall effects and material's resistivity is absent in the weakly insulating range characterized by logarithmic temperature dependence of conductivity. The results provide compelling experimental confirmation to recent models of granular metals predicting transition from logarithmic to exponential conductivity temperature dependence when inter-granular conductance drops below the quantum conductance value and loss of Hall effect scaling when inter-granular conductance is higher than the quantum one. The effect was found at high temperatures and reflects the granular structure of material rather than low temperature quantum corrections.

The extraordinary (or anomalous) Hall effect (EHE) in ferromagnetic materials has attracted much renewed interest. The mechanisms attributed to EHE are divided into two groups: intrinsic and extrinsic. The intrinsic EHE arises in a perfect periodic lattice with spin-orbit coupling due to the topological properties of the Bloch states. The effect does not require any disorder and results in quadratic correlation between the EHE resistivity $\rho_{EHE}$ and longitudinal resistivity $\rho$ ($\rho_{EHE} \propto \rho^2$). The extrinsic EHE mechanisms are due to the asymmetric spin-orbit scattering of spin-polarized electrons on the impurities in the material. One distinguishes between the skew-scattering [1] mechanism for which EHE resistivity scales linearly with longitudinal resistivity ($\rho_{EHE} \propto \rho$) and the side jump mechanism [2] for which the scaling is quadratic ($\rho_{EHE} \propto \rho^2$). When presented using Hall conductivity $\sigma_{EHE} \approx \rho_{EHE}/\rho^2 = \rho_{EHE}\sigma^2$, with $\sigma$ being the longitudinal conductivity, the skew scattering mechanism scales as $\sigma_{EHE} \propto \sigma$ while for the intrinsic and side jump mechanisms $\sigma_{EHE} = const$. Recently, a unified theory of EHE scaling has been proposed for multiband ferromagnetic metals with diluted impurities [3, 4]. The model predicts three distinct scaling regimes in the EHE that are functions of conductivity. In the clean regime ($\sigma > 10^6 \, (\Omega cm)^{-1}$), the skew scattering mechanism is predicted to dominate. The intrinsic contribution becomes dominant in the intermediate disorder regime ($\sigma \sim 10^4 - 10^6 \, (\Omega cm)^{-1}$). In the high disorder range ($\sigma < 10^4 \, (\Omega cm)^{-1}$) the intrinsic contribution is strongly decayed, resulting in a scaling relation $\sigma_{EHE} \propto \sigma^\gamma$ with $\gamma \sim 1.6$. This theory is based on the use of Bloch wave functions assuming a metallic conduction; hence the result is valid only for ferromagnetic metals in principle. However, similar scaling $\sigma_{EHE} \propto \sigma^\gamma$ with $1.33 \leq \gamma \leq 1.76$ has also been predicted [5] for thermally activated hopping processes like variable range hopping, short-range activation hopping or tunneling influenced by interactions in the Efros-Shklovskii regime. Thus, universal scaling in the form $\sigma_{EHE} \propto \sigma^\gamma$ is anticipated for low conductivity materials regardless of whether their conductivity is metallic or thermally activated.

Most of the theoretical works on EHE considered the cases of infinite homogeneous samples. On the other hand many of realistic highly resistive materials like metal-insulator mixtures or thin ferromagnetic films are granular and inhomogeneous, therefore applicability of homogeneous models to these systems is questionable. The theory of Hall effect in granular metals developed in Refs. [6-8] is an interesting deviation from the above mentioned models since it predicts a loss of scaling between Hall effect (both ordinary and extraordinary) and resistivity. The physical reasoning is simple: granular material is described as a network composed of metallic grains with high intra-granular conductivity interconnected by tunnel junctions with low inter-granular conductivity. The Hall voltage (both ordinary and extraordinary) is assumed to be generated within the grains only and not depending on inter-granular connections as long as tunnel resistance is lower than the quantum resistance $R_Q = h/2e^2$. On the other hand, the overall resistance of the system is dominated by the inter-granular tunnel junctions. Thus, no correlation between Hall effect and resistivity is predicted for the granular array, regardless of whether the scaling is satisfied within grains.

In this paper we report on the study of ordinary and extraordinary Hall effects in granular Ni-SiO$_2$ films across the metal-insulator transition. We identify three types of magnetotransport behavior depending on metal content: metallic, weakly insulating and strongly insulating. In the weakly insulating range both the ordinary and extraordinary Hall effects do not depend on material's longitudinal resistivity in agreement with the granular theory of Efetov et al [6-8].

100 nm thick granular films of Ni-SiO2 were prepared by co-evaporation of Ni and SiO$_2$ on room-temperature GaAs substrates using two independent electron beam guns. The deposition rate and the relative volume concentration of the two components were monitored and controlled by two quartz thickness monitors. Usually, a set of up to twelve samples was deposited simultaneously. The relative concentration of the components varied smoothly due to a shift in the geometrical location of the substrate relative to the evaporation sources. Atomic concentration of Ni and SiO$_2$ was measured by energy dispersive X-ray spectroscopy (EDS). The size of Ni crystallites is 3-4 nm and SiO$_2$ matrix is amorphous although there are indications of at least partial crystallization.

Accurate determination of the Hall signal in highly resistive materials is difficult due to a low signal to noise ratio and a strong unavoidable parasitic Ohmic signal. In this work we used the reversed magnetic field reciprocity (RMFR) protocol to extract the Hall signal. According to the RMFR theorem [9-11] switching between pairs of current and voltage leads in a four probe transport measurements is equivalent to reversal of field polarity or magnetization in magnetic materials: $V_{ab,cd}(M) = V_{cd,ab}(-M)$ where the first pair of indices indicates the current leads and the second the voltage leads. The odd in magnetic field Hall term can be separated from a parasitic Ohmic signal by making two measurements at a given field with switched current and voltage pairs and calculating the Hall voltage as: $V_H(B) = 1/2(V_{ab,cd} - V_{cd,ab})$. By using the protocol we succeeded in measuring Hall signal in films with very high resistivity up to 5 Ωcm. An additional important experimental detail is use of high magnetic field. Granular ferromagnets below percolation threshold are superparamagnetic, their magnetization at room temperature reaches saturation at high magnetic fields of few up to 10 Tesla. These samples were measured in fields up to 14 T while the ordinary Hall coefficient was determined in the range 10 - 14 T.

At high magnetic fields when magnetization is saturated perpendicular to the film plane, the Hall resistivity can be presented as: $\rho_{xy} = \rho_{EHE} + R_{OHE}B$, where $\rho_{EHE}$ is the saturated extraordinary Hall effect resistivity, $R_{OHE}$ is the ordinary Hall effect coefficient and $B$ is magnetic field induction. In the following we define $R_{OHE}$ as a slope of the measured $\rho_{xy}(B)$ in the high field range where magnetization is saturated and the slope is strictly linear. $\rho_{EHE}$ was defined by an extrapolation of the linear high field slope at zero field.

Fig.1 presents the saturated extraordinary Hall effect resistivity $\rho_{EHE}$ measured in a series of $Ni_x$-$(SiO_2)_{1-x}$ samples with Ni volume concentration $x$ ranging from 0.72 down to 0.2. The data collected at room temperature and at 77K are shown in two presentations: (a) $\rho$ presentation, in which $\rho_{EHE}$ is shown as a function of resistivity; and (b) $\sigma$ presentation, in which $\sigma_{EHE}$ is given as a function of material's conductivity. Resistivity (and respectively conductivity) of the studied samples change with Ni content by almost five orders of magnitude between $10^{-4}$ to 10 $\Omega cm$. EHE conductivity $\sigma_{EHE}$ spans over 7 orders of magnitude (Fig. 1b) due to normalization of $\rho_{EHE}$ by longitudinal resistivity ($\sigma_{EHE} = \rho_{EHE}/\rho^2$). Overall, $\sigma_{EHE}$ seems to be well described by the ratio $\sigma_{EHE} \propto \sigma^\gamma$ with $\gamma \approx 1.6$ matching the unified scaling prediction [3,4]. However, much more details are disclosed when the same data are shown in the $\rho$- presentation (Fig. 1a). Here three ranges can be identified: $\rho_{EHE}$ increases with increasing resistivity below $10^{-2}\Omega cm$; it is constant for samples in the resistivity range $10^{-2}\Omega cm < \rho < 1$ $\Omega cm$ (Ni concentration range 0.28 > x > 0.22) and grows again when resistivity exceeds 1 $\Omega cm$. In the "low" resistivity range ($\rho < 10^{-2}\Omega cm$) $\rho_{EHE}$ can be fitted to the form $\rho_{EHE} \propto \rho^n$ with the power index $n \approx$ 0.6 at room temperature and $n \approx 0.7$ at 77K. This result is similar to the power law scaling found by Pakhomov et al [12], however it is quite different from the value 0.4 predicted by the unified scaling theory for this range of resistivity. Highly resistive samples with $\rho > 1\Omega cm$ do not follow the power law scaling at all. Instead the data can be fitted by an exponential correlation: $\rho_{EHE} \propto exp(\rho^{1/3})$. Our major focus however is on the plateau range, where EHE resistivity is independent of longitudinal resistivity. Worthy to note that all details of $\rho_{EHE}$ vs $\rho$ variation cannot be traced when the same data are presented in the $\sigma$ - presentation (Fig.1b).

The three ranges identified in $\rho_{EHE}$ vs $\rho$ data can be better understood when we analyze the temperature dependence of resistivity. Fig. 2a presents resistivity as a function of temperature for several samples with different Ni content. The curves are normalized at 77K. Temperature coefficient of resistivity defined as $\alpha = d\rho/dT$ was calculated in the temperature range 77K < T < 85 K. Samples with $\rho < 10^{-2}\Omega cm$ demonstrate metallic-like behavior with $\alpha > 0$, while samples with $\rho > 10^{-2}\Omega cm$ are insulator-like with $\alpha < 0$. $\alpha$ crosses zero in the sample with $x = 0.28$ located at the onset of the $\rho_{EHE}$ plateau. Transition between the metallic-like to insulator-like temperature dependent resistivity in granular matter is usually attributed to the percolation threshold when an infinite metallic cluster is interrupted by an insulating gap. However, granular materials below geometrical percolation can also have a metal-like behavior at high temperatures if the inter-granular tunneling conductance is higher than the intra-granular one: $\sigma_T > \sigma_G$ and the observed $\alpha$ is due to the intra-granular metallicity. In the opposite range with $\sigma_G > \sigma_T$ two cases should be considered: weakly insulating and strongly insulating. Following Efetov and Tschersich [13, 14] the weakly insulating regime occurs in granular systems with inter-granular tunneling conductivity exceeding the quantum conductivity $\sigma_T > \sigma_Q (= \frac{2e^2}{h})$ and is characterized by logarithmic dependence of conductivity on temperature:

$$\sigma = \sigma_0(1 + AlnT) \qquad (1)$$

both in 2D and 3D materials. The strongly insulating range is ascribed to tunnel junction conductivities smaller than the quantum one and is characterized by an exponential variation of conductivity with temperature as:

$$\sigma = \sigma_0 exp[(-B/T)^n] \qquad (2)$$

where n can be different from 1 due to e.g. distribution of grain sizes. It is important to emphasize that the meaning of weakly and strongly insulating granular systems in this context is different from the usually accepted terminology of weak and strong localization. The model [13, 14] is calculated for temperatures high enough to suppress all weak localization effects.

Fig. 2b presents conductivity of samples belonging to the "plateau" range as a function of temperature (in logarithmic scale). The data can be well fitted by the logarithmic dependence in accordance with Eq.(1). Fig. 2c presents the temperature dependence of resistivity for samples beyond the plateau. Resistivity of these samples diverges exponentially as $\rho \propto \exp\left[(T_0/T)^{1/4}\right]$, which is consistent with Eq.(2) for the strongly insulating range [15].

An idealized model of granular material presented in Refs. [6,8,13] can be adapted to more realistic materials by replacing a network of identical spherical grains by finite clusters built of metallic crystallites. We can then use a classical percolation theory to estimate an effective size of clusters as a function of metal concentration in the vicinity of the percolation threshold, and calculate an average resistance of inter-granular tunneling junctions as: $R_T = R_\square t/\xi = \rho/\xi$, where $R_\square$ is sheet resistance, $t$ thickness and $\xi$ is the correlation length or a mean cluster size. The latter can be calculated below the percolation threshold $x_c$ as $\xi = a|x - x_c|^{-\nu}$ with $a$ being the diameter of Ni crystallites and index $\nu$ = 0.88 in 3D systems [16, 17]. By taking $x_c = 0.28$ and $a = 3$ nm we calculated a mean cluster size and a mean inter-cluster resistance for samples below percolation. The arrow in Fig.1a indicates resistivity of material for which the mean inter-granular (inter-cluster) junction resistance equals $R_Q$, which appears to be close to the end of the "plateau" range. We can, therefore identify all samples as belonging to one of three groups: samples with Ni concentration x > 0.28 ($\rho < 10^{-2}\Omega cm$) are metallic, meaning they are either above geometrical percolation or are granular with $\sigma_T > \sigma_G$, the plateau range samples are weakly insulating with $\sigma_G > \sigma_T > \sigma_Q$, and the high resistivity samples beyond the plateau are strongly insulating with $\sigma_T < \sigma_Q$. $\rho_{EHE}$ is thus independent of resistivity in the weakly insulating range but scales with resistivity in metallic and strongly insulating ranges.

The ordinary Hall effect is another subject of our interest. Following Kharitonov and Efetov [6,7] the Hall transport in granular systems, where $\sigma_G > \sigma_T > \sigma_Q$, is essentially determined by the intra-grain electron dynamics. The ordinary Hall resistivity depends neither on the tunnelling conductance $\sigma_T$ nor on the intra-granular mean free path and is given by the classical formula

$$\rho_{OHE} = B/n^*e \qquad (3)$$

where $n^*$ is an effective carrier density inside the grains. Similar to the EHE, the ordinary Hall coefficient is predicted to be independent of material's longitudinal resistivity as the latter is dominated by the inter-granular conductance. We show in Fig. 3 the absolute values of both the ordinary Hall effect coefficient $R_{OHE} = 1/n^*e$ and the EHE resistivity $\rho_{EHE}$ as a function of resistivity. Both parameters demonstrate the same behaviour: they both grow significantly in metallic samples with increasing resistivity, saturate simultaneously in the "plateau" range, and change again in the strongly insulating range. $R_{OHE}$ of the sample with the highest resistivity we succeeded to measure beyond the plateau is multiplied by a factor -0.5, as marked in the figure. Remarkably, polarity of $R_{OHE}$ reverses from negative in metallic and weakly insulating ranges to positive in the strongly insulating one. This feature is beyond the scope of this paper and we shall only note that similar change of polarity across the percolation threshold has been reported in W-Al$_2$O$_3$ [18].

$R_{OHE}$ as a function of $\rho_{EHE}$ is shown in the inset of Fig.3 for samples in the metallic and weakly insulating ranges. The correlation between the two parameters is linear, which is intriguing since it implies that $\rho_{EHE}$ is inversely proportional to an effective density of carriers available for diffusive metallic conductivity. We are aware of only one model in which a straightforward correlation between EHE current and carrier density has been calculated [19]. Applicability of this mechanism to granular metallic ferromagnets should be reconsidered.

Saturation of EHE resistivity in the vicinity of percolation threshold has been reported in ultrathin films of CNi$_3$ [20] and FePt [21]. In both cases the phenomenon was observed at low temperatures, and interpreted as an evidence of quantum corrections to EHE. Also, saturation of the ordinary Hall coefficient was observed in non-magnetic Cu-SiO$_2$ [22]. This saturation and orders of magnitude enhanced values of $R_{OHE}$ were interpreted in terms of quantum percolation. The results reported here were obtained at room temperature and at 77K and are therefore not related to low temperature quantum phenomena.

In summary, two qualitative properties have been theoretically predicted for granular metals depending on inter-granular tunnelling conductance: transition from logarithmic to exponential temperature dependence of conductivity when inter-granular conductance drops below the quantum conductance value [13, 14], and loss of scaling between both the ordinary and extraordinary Hall effects and material's resistivity when inter-granular conductance is higher than the quantum one [6-8]. We found experimental confirmation to both these predictions in granular ferromagnetic Ni-SiO$_2$ mixtures. In particular, scaling of the extraordinary Hall effect with resistivity is absent in granular structures when Hall effect is generated by diffusive scattering within grains and the total resistivity is dominated by inter-granular tunneling. The effect was found at high temperatures and reflects the granular structure of material rather than low temperature quantum corrections. We also found a linear correlation between the EHE resistivity and the ordinary Hall coefficient implying a straightforward dependence of EHE on an effective density of carriers.

**Figure captions.**

Fig.1. (a) The saturated extraordinary Hall effect resistivity $\rho_{EHE}$ measured in a series of $Ni_x$-$(SiO_2)_{1-x}$ samples as a function of resistivity at room temperature (solid circles) and at 77 K (open circles). Ni volume concentration $x$ ranges from 0.72 down to 0.2. The arrow indicates resistivity of material for which the mean inter-granular (inter-cluster) junction resistance equals quantum resistance $R_Q$ (see text).

(b) The same data in $\sigma$ presentation, in which $\sigma_{EHE}$ is shown as a function of material's conductivity.

Fig.2. (a) Resistivity of a series of $Ni_x$-$(SiO_2)_{1-x}$ samples as a function of temperature. The data are normalized at 77 K. Indices indicate Ni concentration.

(b) Normalized conductivity of two "plateau" range samples as a function of logarithm of temperature.

(c) Logarithm of resistivity of "beyond the plateau" samples as a function of $T^{-1/4}$.

Fig.3. Absolute values of the room temperature ordinary Hall effect coefficient $R_{OHE}$ and the EHE resistivity $\rho_{EHE}$ as a function of resistivity. $R_{OHE}$ of the highest resistivity sample beyond the plateau is multiplied by a factor -0.5. Inset: $R_{OHE}$ as a function of $\rho_{EHE}$ for samples with resistivity below 1Ωcm. Straight line is guide to the eye.

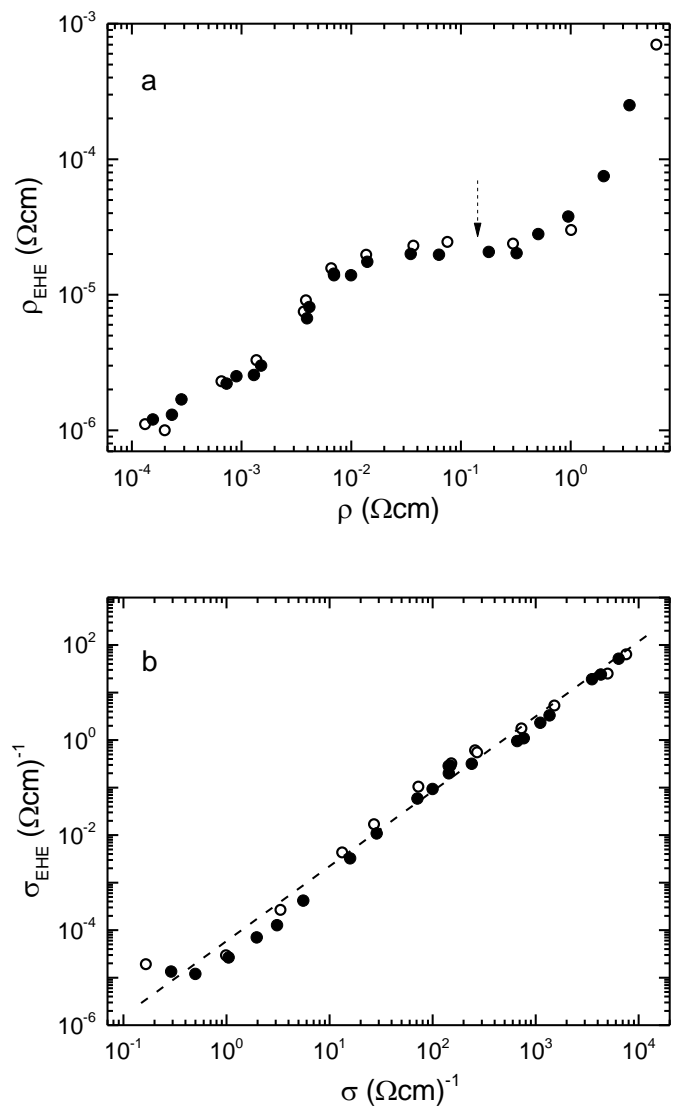

Fig. 1

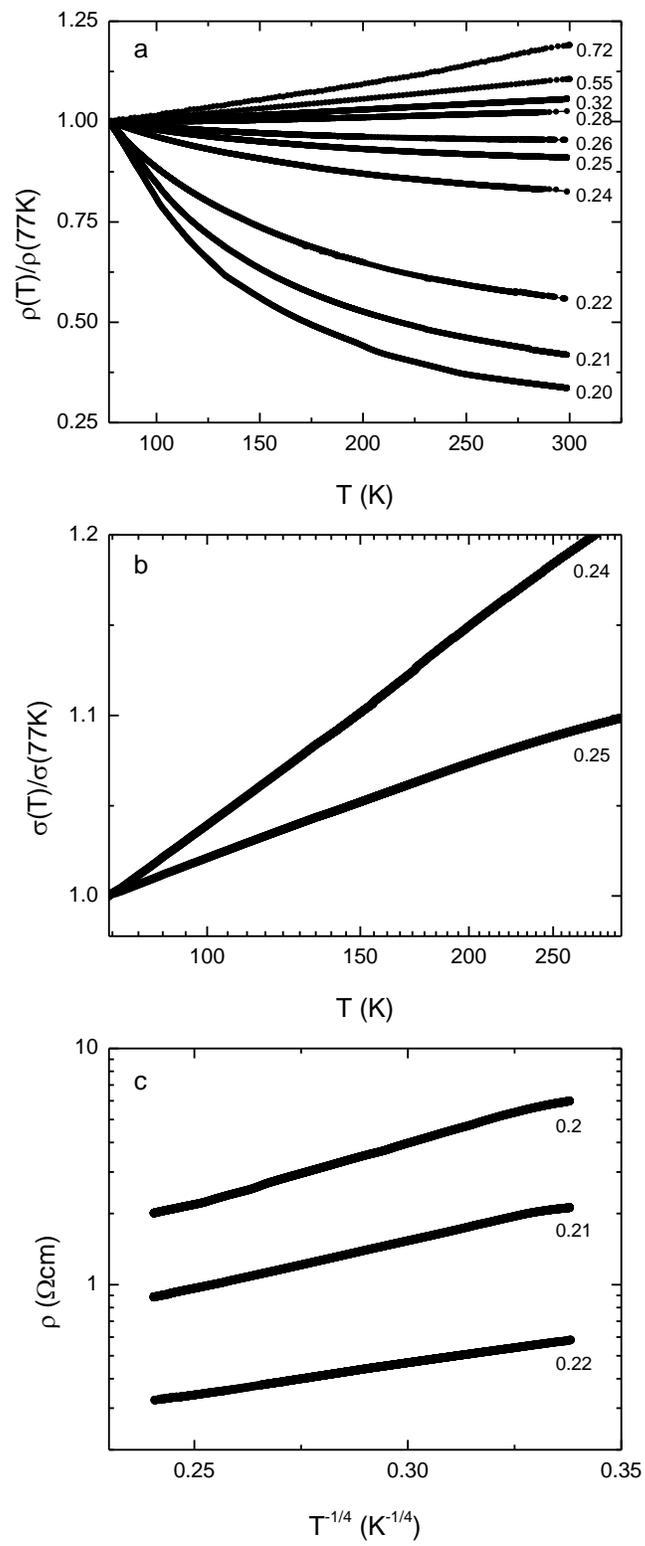

Fig. 2

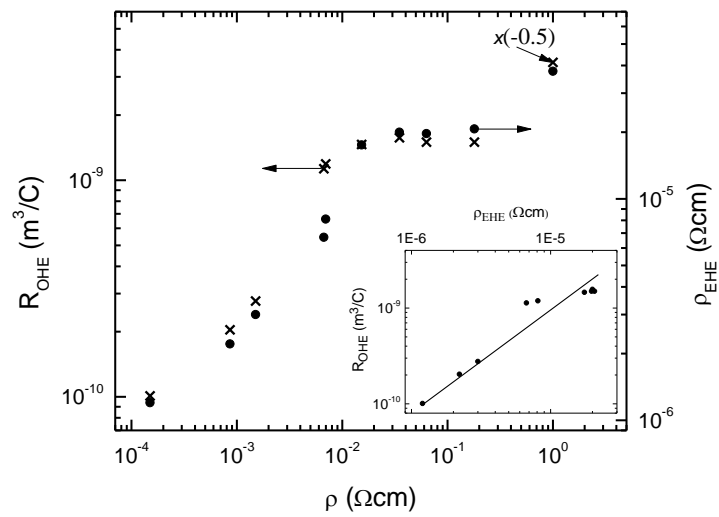

Fig. 3